\documentclass[conference]{IEEEtran}
\IEEEoverridecommandlockouts
\usepackage{cite}
\usepackage{amsmath,amssymb,amsfonts}
\usepackage{algorithmic}
\usepackage{graphicx}
\usepackage{textcomp}
\usepackage{xcolor}
\usepackage[numbers]{natbib}

\usepackage{etoolbox}
\makeatletter
\pretocmd{\NAT@citexnum}{\@ifnum{\NAT@ctype>\z@}{\let\NAT@hyper@\relax}{}}{}{}
\makeatother

\def\BibTeX{{\rm B\kern-.05em{\sc i\kern-.025em b}\kern-.08em
    T\kern-.1667em\lower.7ex\hbox{E}\kern-.125emX}}
\begin{document}

\newcommand{\comment}[1]{}
\title{Catch Me If You GAN: Using Artificial Intelligence for Fake Log Generation
}

\author{\IEEEauthorblockN{1\textsuperscript{st} Christian Tömmel}
\IEEEauthorblockA{\textit{Institute for Wireless Communication and Navigation} \\
\textit{University of Kaiserslautern}\\
Kaiserslautern, Germany \\
toemmel@eit.uni-kl.de}
}

\maketitle

\begin{abstract}
With artificial intelligence (AI) becoming relevant in various parts of everyday life, other technologies are already widely influenced by the new way of handling large amounts of data. Although widespread already, AI has had only punctual influences on the cybersecurity field specifically. Many techniques and technologies used by cybersecurity experts function through manual labor and barely draw on automation, e.g., logs are often reviewed manually by system admins for potentially malicious keywords.

This work evaluates the use of a special type of AI called generative adversarial networks (GANs) for log generation. More precisely, three different generative adversarial networks, SeqGAN, MaliGAN, and CoT, are reviewed in this research regarding their performance, focusing on generating new logs as a means of deceiving system admins for red teams. Although static generators for fake logs have been around for a while, their produces are usually easy to reveal as such. Using AI as an approach to this problem has not been widely researched. Identified challenges consist of formatting, dates and times, and overall consistency. Summing up the results, GANs seem not to be a good fit for generating fake logs. Their capability to detect fake logs, however, might be of use in practical scenarios.

\end{abstract}


\section{Introduction}\label{section:introduction}

Ever since the internet's first cornerstones in the 70's, the Arpanet, it has been growing in different directions and introducing diverse use cases.
Naturally, this growth also brought financial interests and made internet services a target of more and more sophisticated attacks.
With such a large ecosystem at stake, computer professionals in the cybersecurity domain seem to play a never-ending game of cat and mouse with malicious parties.
Many building bricks to security were introduced in the last decades, and some of the latest advancements, one of these being artificial intelligence (AI), introduced entirely new possibilities to defend, but also to attack security perimeters.

In its recent days, AI has proven to be helpful regarding a diverse set of challenges:
image generation \cite{stylegan2}, optimization of forecasts \cite{forecasting}, speech synthesis \cite{speechsynthesis}, and early diagnoses in medicine \cite{earlydiagnosis}, are just a few examples. However, some of these fantastic use cases can also be adapted to create outcomes which can be leveraged to serve malicious purposes.

A certain website called "This Person Does Not Exist"\footnote{
https://thispersondoesnotexist.com} depicts just one of these use cases and has drawn a great deal of attention from the general public to the possibilities, but also to the dangers of AI.
The website's creators use a generative adversarial network (GAN) to produce images of real-looking faces based on a dataset which the GAN's generator and discriminator are trained with.
Although this poses only a single use case, the implied possibilities using this technique are wide-ranging.
It has become easy to create fake profiles using randomly generated imagery, and the tools are becoming better and better. \citeauthor{dfgendet} have researched this in depth in \cite{dfgendet}.

The latest publicly known progressions in this field, so called "Deepfakes", are even capable of creating video sequences of people who are moving and talking. For example, a celebrity's head might be swapped with the head of someone else in a recorded video. According to studies, these scenes look realistic enough to deceive most people of them being AI, and false acceptance rate (FAR) is as high as roughly 75 percent as found by \citeauthor{korshunov2020deepfake} in \cite{korshunov2020deepfake}. This enables putting virtually every desired word in someone's mouth, simply by collecting base data about their looks, i.e., images, and voice, i.e., audio recordings.  
The implied concerns for security and privacy of this recent breakthroughs are striking: virtually everyone with enough content of them on the internet can be used as baseline data to create an artificially generated copy of them speaking, moving and acting as favored by the content creator.
This takes away the credibility of basically every content regarding this person on the internet as long as it is not explicitly confirmed by a trusted source of truth.
Threats with respect to the aforementioned possibilities may even be of national concern \cite{Nguyen.25.09.2019}.

Although AI introduces previously nonexistent risks to private persons, companies and even nations, it can also be leveraged to enhance security.
Especially GANs yielded useful results for diverse tasks in the past. For example, in defensive security. GANs are useful for tasks such as obscuring sensitive information \cite{obscure}, malware detection \cite{ransomwaredetection}, steganography \cite{steganography}, or as a new field of cryptography – neural cryptography \cite{neuralcrypto}.
However, various other use cases in offensive security tend to be overlooked. Defensive security as the classical branch of security is researched more deeply with respect to AI, yet a variety of potentially interesting topics have not been researched deeply as of now. 

With this contribution, the possibilities of GANs for fake log detection and generation shall be researched. Three different GANs were selected and evaluated regarding their use for fake log generation based on a dataset consisting of Windows log entries.
The remainder of this work is structured as follows. Section \ref{section:2} gives an overview on the background, introducing both offensive security as well as GANs. In Section \ref{section:3}, the test setup, execution and findings are described. Section \ref{section:4} gives a discussion on the findings and their relevance for the future. The last section, Section \ref{section:5}, concludes this contribution and shows possible future work.

\section{Background}\label{section:2}

\subsection{Generative Adversarial Networks}
The term "Generative Adversarial Network" denominates a form of unsupervised learning.
As proposed by \citeauthor{Goodfellow.10.06.2014} in \cite{Goodfellow.10.06.2014}, two artificial neural networks are set up to play a zero-sum game. One of the networks, the generator, produces data and the second network, the discriminator, tries to correctly guess whether the data it receives is artificially generated or not, i.e., actual data.

More precisely, the generator G is trained to minimize the probability of the discriminator D assigning the correct label. At the same time, D is trained to maximize the likelihood of assigning correct labels to both, actual input data as well as G's generated information. With $p_z(z)$ representing noise and $p_g$ representing the generator's data, two neural networks $D$ and $G$ are considered where $D$ outputs a single scalar which is the discriminators assigned label to a data point. This yields a minimax game with value function $V(D, G)$.

\begin{equation*}
\begin{aligned}
\min_G \max_D V(D,G) &= \\
&\mathbb{E}_{x \sim p_{data}(x)} [log D(x)] + \\
&\mathbb{E}_{z \sim p_z (z)} [log(1-D(G(z)))]
\end{aligned}
\end{equation*}
\\[-0.6em]

The discriminator is trained on distinguishing which data comes from real samples and which data was generated by the generator. The above equation from \cite{Goodfellow.10.06.2014} shows the eponymous adversarial approach where two neural networks with contrary goals each try to win the zero-sum game and deceive, or not get deceived by respectively, the other one.

GANs have performed well in tasks like image generation and speech synthesis, but have only a few use cases in cybersecurity as surveyed by \citeauthor{Dutta.1028202010312020} in \cite{Dutta.1028202010312020}. To understand what cybersecurity refers to in the context of this work, the next subsection will explain the background.

\subsection{Defensive and Offensive Security}
Within the term cybersecurity, in general two fields are considered during enhancement of overall security: defensive security and offensive security \cite{offdef}. They partly depend on each other and foster each other's efficiency, so a strict distinction might not always be suitable; however, the overall goal of both is to jointly discover and fix weak points in a given architecture or organization. For example, a penetration test might be conducted which then finds vulnerabilities. These findings can then be dealt with in defensive ways, e.g., by analyzing the source code of the vulnerable parts and applying patches for identified weak points. Source code analysis with respect to security is, by itself, an entire field of research which uses methods like static code analysis \cite{staticcodeanalysis}.

In the field of defensive security, the aim is to be on the defender's side and try to keep out malicious parties. This can happen in various ways, primarily either proactive or reactive. Whilst a proactive approach is set to not let unauthorized individuals in, reactive approaches consist of measures which remove those attackers after they initially gained access or tried to gain it \cite{reacproac}.

Defensive security primarily relies on prevention and detection of attacks as well as reactive measures.
Vulnerabilities are recognized and fixed accordingly, ideally before being exploited. Approaches like "Defense-in-Depth", were multiple security mechanisms are layered to strengthen the security level, are widely used, for example with industrial control systems as discussed by \citeauthor{didict} in \cite{didict}.

Offensive security, however, is not focused on patching and fixing errors in the first place, but on putting individuals – so called "penetration testers" – in the shoes of a malicious party and having them attack a given application, infrastructure or organization. Through mimicking of attacks, vulnerabilities can be found and then fixed which could be exploited by actually malicious attackers if left undiscovered \cite{offsec}.

An overview on the ecosystem of offensive security tools was given by \citeauthor{DuqueAnton.2020} in \cite{DuqueAnton.2020}. The authors collected and analyzed 35 tools, assigning them to the different phases of the MITRE ATT\&CK matrix.

With respect to this research topic, the status quo for fake log generations is, or more generally, functions similar to flog\footnote{https://github.com/mingrammer/flog}. Flog supports a number of services while functioning static in nature. The general approach needs to be adapted and implemented for every desired log format. Conventional techniques rely on static concatenation of mostly randomly generated strings where each substring fulfills the expectations on this very part of the log entry. Although this yields syntactically valid log entries, the entire log file generated with this approach is fundamentally flawed. Multiple samples do not build up on one another and there is no coherence overall as many parts and generated randomly. Someone analyzing these forged entries can easily spot their lack of coherence. 

\subsection{AI in Security}

There is a vast amount of use cases for AI in defensive security such as AI-based intrusion detection systems (IDS) as proposed by \citeauthor{aiids} in \cite{aiids}, or within the realm of cryptography in form of neural cryptography as a measure of encryption \cite{Pattanayak.2018} as proposed by \citeauthor{Pattanayak.2018}. The area of AI in defensive security has already been highlighted by many others, hence, the focus of this research is AI in offensive security.

Various tools used by penetration testers function as password, or more general, term guessers. Tools like johntheripper and hashcat can be used to brute force passwords, i.e., trying all possible combinations of letters, numbers, etc. for all the positions within different lengths of a potential password until a correct one is found. 
One of thje more sophisticated approaches stems from the authors \citeauthor{hitaj2019passgan} They described a new approach based on GANs in \cite{hitaj2019passgan}. Their overall goal is to use GANs to generate better password brute force lists. While tools like hashcat allow for patterns or masks to enrich a password list, these static approaches only bring a limited amount of profit. PassGAN generates entirely new passwords which improve brute forcing. Other approaches based on AI, specifically ones using recurrent neural networks or Markov chains, seem to be capable of similar augmentations to password lists \cite{rnnmarkov}, although there exists no scientific comparison of these approaches.

Another example of AI in offensive security is DeepExploit\footnote{From https://github.com/13o-bbr-bbq/machine\_learning\_security}. DeepExploit uses the A3C machine learning model in combination with metasploit to learn to attack systems and even generate a report out of it. The steps consist of the classical path a penetration tester would follow, starting with enumeration, exploitation as well as post exploitation to gather more information for its user to conduct privilege escalation. It can then spread to other systems.

Besides some implementations like DeepExploit, there exists only limited research regarding AI use cases for attackers. Although approaches like DeepExploit show how AI can be helpful in a red/blue team scenario, there are various other scenarios where AI might be of use for a malicious party.

One use case which could be particularly interesting for both attackers and defenders is fake log generation and detection, respectively.

\section{Generating Logs with Generative Adversarial Networks}\label{section:3}

The generation of fake logs and detection, likewise, poses a use case which can be interesting for both parties – attackers and defenders. Red teams, for example, might want to place fake logs to deceive the opposing blue team and stay undetected. Blue teams, however, might profit from a fast way to analyze logs not only for statically identified anomalies, but also regarding deeply interwoven contradictions, potentially hundreds of single entries apart. The latter would be a functionality of the discriminator while the first is the generator's task. In contrary to the static approach of flog, as mentioned earlier, an approach based on GANs could help with more easily generating better quality log entries and possibly switching formats.

In order to generate log entries and evaluate how effective in terms of deception a generated log is, it must first be discussed which building blocks make up a single log entry and how it relates to other log entries. The first subsection dives into the properties of log entries and answers this question whereas the following two subsections deal with the setup of GANs for fake log generation and the results.

\subsection{Properties of Log Entries}
Although the specific details of log entries are strongly dependent on the application which produces logs, some rather basic and general properties which should be existent in one form or another in all logs can be identified. 
In the following, log properties are differentiated as being either syntactic or semantic.

Syntactic properties are those which describe what a single log entry is made up of, how individual parts are delimited from another and what each of these parts describes. The identified syntactic properties are summarized in Table~\ref{table:synprop}.

\begin{table}[htbp]
\caption{Syntactic properties of log entries}
\centering
\begin{tabular}{|| c ||c c||} 
 \hline
 Property & Example value & Content \\ [0.5ex] 
 \hline\hline
 Date and time & 2021-09-30T14:00:07 & Current time$^{\mathrm{a}}$ \\ 
 \hline
 Event identifier & EFR (error: file read) & ID for an event\\
 \hline
 Event description & Could not access file & Description of event  \\
 \hline
\multicolumn{3}{l}{$^{\mathrm{a}}$E.g., compliant to ISO 8601, or a timestamp}
\end{tabular}
\label{table:synprop}
\end{table}

Depending on rather cosmetic formatting decisions, an example log entry following the above property structure might look as follows when using whitespace separation of individual fields:

\begin{equation}
\texttt{20210830T104958~~EFW~~Write failed}
\end{equation}
\\ [-3.0ex]

The above example shows a failed file write operation without naming a specific file using ISO 8601 date and time notation\footnote{https://www.iso.org/iso-8601-date-and-time-format.html} as listed in the first field. EFW could be any (short) code for a certain class of events. It could also be a more general scheme, simply using "error", "warning", and "success" for the distinction between event types. In this particular case, EFW refers to an \textbf{e}rror which occurred during a \textbf{f}ile \textbf{w}rite operation. These short codes are often specific to an application.

Usually, properties are, depending on the software used, differentiated in a structured way within a single log entry. Whilst there exist web APIs for log output which adhere to JavaScript Object Notation (JSON) formatting, Windows does use a whitespace separation in its logs to differentiate between properties. 

Depending on the application, some logs might also contain additional properties. Some logs reviewed have shown an event type like "success", "error" or "warning". Logs of web applications might also contain the IP address which was involved in a logged action, whereas logs of an operating system might list related processes for a given log entry. The goal here is to analyze log entries in their most basic form, hence, only these three syntactic properties are reviewed.

Semantic properties of log entries, on the other hand, are defined by the individual entries' actual meaning and its intercorrelation to other entries. In general, some events can only occur after a certain other event, e.g., the crash of a process cannot happen before it is launched. And these event chains might not even span over multiple events in a row, but over a large number of events with multiple different listings in between. A real log file will always have this property fulfilled as it is just the natural way of logs being created and a process cannot crash before it has been ordered to start, but it requires contextual awareness and rather complex memory functionalities for AI to correctly recreate this, potentially spanning over gigabytes of log files. Table~\ref{table:semprop} summarizes the semantic properties.

\begin{table}[htbp]
\caption{Semantic properties of log entries}
\centering
\begin{tabular}{|| c ||c||} 
 \hline
 Property & Description \\ [0.5ex] 
 \hline\hline
 Chronology & Entries should follow an overall chronology  \\ 
 \hline
 Event coherence & Dependent events can only happen in one order \\
 \hline
\end{tabular}
\label{table:semprop}
\end{table}

On the most basic level, log entries should follow at least two properties. Firstly, they need to be in the correct order. New log entries are either appended at the end or at the top of an existing log file, hence, the times or timestamps for log entries are either monotonically increasing or decreasing. As some events might, depending on the time precision used, happen at the same time, this property does not need to be strict. This first property is called chronology. Secondly, the events must happen in a coherent order. It is not possible for a process to write to a file if that process has never been started, hence, the start of the process is always earlier in time than the write action of said process. This means that all events are coherent.

The question that comes up is whether GANs can learn these properties and generate fake data which
\begin{itemize}
\item adheres to the syntactical and semantical properties, and
\item can be used to deceive an experienced human evaluator.
\end{itemize}
In order to answer this question, some practical tests for three different types of GANs for log sample data as described in the following were conducted.

\subsection{Setup}

Three components are relevant for the setup: the training and test data and its source, the GANs used for comparison, and the hardware used to run the tests. 

Regarding data, a common Windows log dataset by zenodo\footnote{https://zenodo.org/record/3227177} is used during the experiments. It offers roughly 28 gigabytes of "Component Based Servicing" (CBS) log data. In addition to its size, other pluses for this dataset are the clear structure of individual log entries, whitespace separation and almost ISO 8601 compliant times, only off by using a whitespace instead of capital T. This time format is static in its form and can potentially be learned as easily. For training data, 50.000 random log entries of the dataset were used, whilst another 20.000 random, but different entries function as testing data.

To gain broader insights on the usefulness of GANs, GAN implementations based on williamSYSU's approaches\footnote{https://github.com/williamSYSU/TextGAN-PyTorch/} were used as a basis. The author of these has provided ten different GAN implementations, based on their respective research, which can be configured easily, but can also be used in default mode.

In the experiments, the first GAN used is Sequence Generative Adversarial Network (SeqGAN), as proposed by \citeauthor{seqgan} in \cite{seqgan}. SeqGAN is a certain type of GAN which aims to smooth away difficulties that general GANs have, e.g., passing the gradient update from the discriminator to the generator, or balance a score for not fully generated text sequences \cite{seqgan}. The second GAN is Maximum-Likelihood Augmented Discrete Generative Adversarial Network (MaliGAN) as introduced by \citeauthor{maligan} in \cite{maligan}. MaliGAN specifically uses a low-variance objective generated from the discriminator's output which functions like a back propagation in order to be able to deliver contextual awareness and remember past events. The third GAN is called Cooperative Training (CoT), which was initially described by \citeauthor{cot} in \cite{cot}. CoT uses the minimax game of GANs, and transforms it into a joint maximization problem.

For all GANs – SeqGAN, MaliGAN, and CoT – a configuration deviating from the proposal of their respective introducers was used, calculating 50 maximum likelihood estimator (MLE) epochs and 3 adversary epochs. 

Regarding training, a personal computer using an 11th gen Intel i9 CPU in combination with 32gb of DDR4-RAM and an NVIDIA RTX 3080 was used.

\subsection{Results}
After training, the results in terms of sample data as well as accuracy can be reviewed.
The accuracies among SeqGAN, MaliGAN, and CoT as well als generator G and discriminator D losses and negative log-likelihoods (NLL) are listed in Table~\ref{table:trainresults}.

\begin{table}[htbp]
\caption{Training results}
\centering
\begin{tabular}{|| c ||c c c c c||} 
 \hline
 GAN & G loss & D loss & G NLL & D NLL & Acc \\ [0.5ex] 
 \hline\hline
 SeqGAN & 201.33 & 0.29 & 0.12 & 0.07 & 0.89 \\ 
 \hline
 MaliGAN & 12.92 & 0.31 & 0.11 & 0.08 & 0.88 \\
 \hline
  CoT & 8.51 & 8.59 & 0.13 & 0.09 & 0.83 \\
 \hline
\end{tabular}
\label{table:trainresults}
\end{table}

The NLL of each GAN is on par with the other GANs. The accuracy of CoT is visibly below the accuracy of SeqGAN and MaliGAN. Only for the generator and discriminator losses, large differences could be identified.
Regarding properties, the syntactic properties were mostly mimicked in a correct way by the GANs after manual review of sample data as shown in Table~\ref{table:synresults}.

\begin{table}[htbp]
\caption{Results for syntactic properties}
\centering
\begin{tabular}{|| c ||c c c||} 
 \hline
 \# & Date and time & Event identifier & Event description \\ [0.5ex] 
 \hline\hline
 SeqGAN & \checkmark & \checkmark & \checkmark\\ 
 \hline
 MaliGAN & \checkmark & \checkmark  & \checkmark\\
 \hline
 CoT & \checkmark & x & x \\
 \hline
\end{tabular}
\label{table:synresults}
\end{table}

For semantic properties, none of the GANs were able to adhere to chronology and event coherence as shown in Table~\ref{table:semresults}.

\begin{table}[htbp]
\caption{Results for semantic properties}
\centering
\begin{tabular}{|| c ||c c||} 
 \hline
 \# & Chronology & Event coherence \\ [0.5ex] 
 \hline\hline
 SeqGAN & x & x \\ 
 \hline
 MaliGAN & x & x \\
 \hline
 CoT & x & x \\
 \hline
\end{tabular}
\label{table:semresults}
\end{table}

Review of actual samples that were generated during the experiments shows that they do not resemble the training data close enough to deceive security analysts with respect to the test setup. The generator's samples have a large amount of whitespaces in different places for all the GANs. After statically cleansing the excess whitespace, the log entries look far more deceiving regarding their syntactic properties, but semantic properties are not adhered to.

Also, the dates generated by the GANs are syntactically incorrect in several samples. E.g., too many numbers were present for the data to correctly represent a date. Furthermore, regarding dates, the chronology of log entries is not ensured by any of the GANs. More precisely, the first log entry should have either the oldest or the newest date and all following entries must be either earlier in time or later, respectively. This very basic semantic property of log entries is hard to adhere to by GANs as learning it and feed forward the old date for later samples is not trivially done in most GANs.

All in all, regarding syntactic properties, the three GANs were able to mimic the fundamental structure of single log entries, i.e., starting with a date, followed by an event identifier and an event description. The individual parts of each single log entry are good enough to deceive analysts with, except for a few cases where times and dates were not valid. Nevertheless, they failed to mimic the event coherence and chronology of events. 

\section{Discussion}\label{section:4}

As GANs are highly prone to this kind of problem, overtraining occurred during the training of CoT. The GAN overtrained on whitespace and eventually degraded to only produce dates, times and whitespace. It still gets a good accuracy which shows that one has to be careful when reviewing such values.

The repeating dates and times that appeared are not surprising as the GAN has only received certain date and time combinations and tries to mimic those.
GANs interpret the numbers (i.e., multiple digits) as single symbol instead of different separable units which function individually and can be combined in various ways.
They cannot easily learn this behavior of dates and times. It might make sense to generate times in a static fashion and use GANs for the other parts only. This static date and time generation also has the advantage of trivially staying in line with the chronology property.

We performed experiments with 70.000 data points and few epochs. Although larger training sets and more epochs or generations can help to train a better model, it appears to us that GANs are not quite the best fit the problem of fake log generation, judging from the vast deviations encountered. 

Furthermore, models trained as described above would not be fit to forge logs other than Windows CBS logs. The log extraction and training process would need to be conducted for each application as most logging has arbitrary structures, formats or application-specific terminology.

After reviewing the results and discussing the setup, its strengths and weaknesses, we can conclude that GANs might not be fit for generating or detecting fake log files. There is a vast amount of complexity, potentially spanning over thousands or even millions of individual log entries, and the evaluated GANs were not able to suffice. The generated logs, even if the discriminator could be deceived, would not be able to deceive an experienced professional. Although this is clearly visible from the experiments, we can expect that, using far more potent hardware, it might be possible to build better models. We can still conclude that other types of AI like transformers (e.g., GPT-3) might be better at generating fake logs. 

Although a negative result was observed, further research should be conducted to verify the findings. Possibly, the use of other formats for dates and times might lead to more convincing fake logs. The challenges that were encountered in the above attempts, which partly might be due to the setup, should be addressed. Among these are repeating dates and times which are not in a chronological order, and a large number of whitespaces.
A more complex setup might yield better results, e.g., by using larger sample sizes, more training epochs and potentially even evaluating a wider variety of GANs. 

Reviewing the results, it is unlikely that even junior security analysts can be deceived by fake logs which were generated by the evaluated GANs under the described setup conditions.

\section{Conclusion and Future Work}\label{section:5}

We started with shortly introducing the relevant background information regarding security and AI.

Afterwards, we discussed the properties based on which we can compare generated log entries and identify a single log entry. We continued by reviewing the setup and the different GANs used during the experiments. This step was followed by the examination of the results. We found that the examined GANs – SeqGAN, MaliGAN, and CoT – do not perform well with respect to the aforementioned setup. 

We then discussed the results and concluded that other types of AI could be more suitable for generating or detecting fake logs. Some rather generic tasks, e.g., generating dates and times, can more easily be dealt with by simple and not as sophisticated methods. Potentially, future research might look into the advantages of using other types of AI like long short-term memory approaches or transformers.

Also, other aspects of AI in security are interesting. For example, the field of (semi-)automated penetration testing based on AI is related to deception of security analysts if an automated attacker system tries to cover its tracks.

All in all, GANs did not perform well in the tests. Although only very limited capacity was available for calculations (i.e., few epochs and generations), this first probe indicates that GANs are not the right way for fake log generation. This failure might have been avoided with more complex approaches and further research should be conducted.

\bibliographystyle{IEEEtranN}
\bibliography{bib}
\end{document}